# Computational Materials: III-V Semiconductor Clusters

Hamidreza Simchi[1,2]

[1]Department of Physics, Iran University of Science and Technology, Narmak, Tehran

16844, Iran

[2]Semiconductor Technology Center, Tehran Iran

**Summary:** Semiconductor materials have been playing an important role in daily life. Among them, the narrow gap III-V semiconductor materials InAs, GaSb, InSb and their alloys are particularly interesting and useful materials since they offer the promise of being able to access the 2-10 microns wavelength region and should provide the next generation of LEDs lasers and photo detectors for applications such as sensors, molecular spectroscopy and thermal imaging. On the other hand the discovery of fullerenes and nanotubes of carbon have led to a wide spread interest to understand their properties, applications and development of unconventional forms (i.e Clusters) of materials. Therefore it is expected, that, the clusters of III-V semiconductors have been widely studied. One of the main interests in semiconductor clusters is the variation of band gap with size which affects the photoluminescence properties. Here we review the progress in our understanding of the structure, and electronic spectra of InAs, GaAs, InSb and CdSe clusters. First principles approaches based on Hartree-Fock and Density Functional Theory have become central to such studies and a brief overview of them is given.

**Contents**

# 1. Introduction

The progress made in physics and technology of semiconductors depends mainly on two families of materials: the group IV elements and the III-V compounds. The first report of the formation of III-V compounds was published in 1910 by Thiel and Koelsch [1]. They synthesized a compound of indium and phosphorus and reached the conclusion that its formula is very probably InP. The fact that one of the III-V compounds, InSb is a semiconductor akin to Ge and $\alpha$-Sn, was reported in 1950 by Blum, et al [2]. The particular features that attracted interest were the low effective mass, high electron mobility, and the ionic component in the crystal binding. After the first exploratory research, nearly all of the work was limited to InSb, partly because this was an almost ideal vehicle for such studies, but also because it was easy to grow InSb single crystals. The technology of the other compounds fell far behind this, for the purification was very difficult and there was no great incentive to work in this field. The invention of the semiconductor laser and the discovery of the Gunn effects have caused a marked change in this situation. There is now an increased prospect of industrial application of these materials, and we may expect a growing interest in GaAs, GaP, and compounds or alloys of compounds which can emit visible or microwave radiation. Now a days, III-V compound semiconductors are widely used as substrates for optical devices such as LED's and laser diodes and for electronic devices such as FET's, HEMT's, HBT's and IC's. These applications are becoming key elements in an advanced information society. A complete overview of the technologies necessary to grow bulk signal-crystal substrates, and grow hetero-and homoepitaxial III-V films using molecular beam epitaxy (MBE) or metal-organic chemical vapor deposition (MOCVD) is provided in reference [3].

On the other hand the discovery of fullerenes and nanotubes of carbon have led to a wide spread interest to understand their properties, applications and development of unconventional forms (i.e Clusters) of materials. Like crystals a molecule can be represented as a system of electrons traveling in the field of ions. Hence the approach where the crystal is simulated by a certain "large molecule" fragment with corresponding boundary conditions is quite valid. Such a large molecule has been called a cluster while the approach itself is called the cluster approach. The cluster approach utilizes a variety of techniques from quantum chemistry employed to describe the molecular electronic properties. Of course clusters are not suitable for describing physical phenomena associated with long-range order due to their finite size. At the same time phenomena associated with short-range order can be successfully analyzed by means of the cluster approach. For example, point defects in crystals [4], localized excitations, self-localized quasiparticles, chemisorption [5], etc., have been successfully analyzed within the framework of the cluster approach.

Here we review the progress in our understanding of the structure, and electronic spectra of InAs, GaAs, InSb and CdSe clusters. First principles approaches based on Hartree-Fock and Density Functional Theory have become central to such studies and a brief overview of them is given.

## 2. Computer simulations using first principles approaches

### a. Self-Consistent Field (SCF) Approximation

In considering the motion of electrons in condensed matter we are dealing with the problem of describing the motion of an enormous number of electrons and nuclei (~ $10^{23}$) obeying the laws of quantum mechanics. An exact solution of this problem is impossible and hence it is necessary to rely on a wide range of simplifying approximation.

Consider a system of N interacting electrons which move in an external potential, set up, for example, by the positively charged nuclei of the system. We can divide the Hamiltonian *H* of the electronic system into two parts:

$$H = \sum_{i=1}^{N} h(i) + \frac{1}{2} \sum_{i \neq j} v(i,j) \quad (2.1.1)$$

The single-electron operator

$$h(i) = -\frac{1}{2m}\nabla_i^2 + v(i) \quad (2.1.2)$$

represents the sum of the kinetic energy $-(1/2m)\nabla_i^2$ of the electron and its energy in the external potential V(i). Should the external potential be attributable to nuclei of charges $-Z_\lambda e^2$, we would have $v(i) = -\sum_\lambda Z_\lambda e^2 / r_{\lambda i}$, where $r_{\lambda i}$ is the distance between nucleus λ and electron i. The term $v(i,j) = e^2 / r_{ij}$ describes the Coulomb repulsion two electrons i and j which are a distance $r_{ij} = |r_i - r_j|$ apart.

Next the Hamiltonian is written in second quantized form, for which we introduce electron field operators $\psi(\vec{r})$. They satisfy the following anti-commutation relations:

$$[\psi_\sigma^+(\vec{r}), \psi_{\sigma'}(\vec{r}')]_+ = \delta_{\sigma\sigma'}\delta(\vec{r}-\vec{r}')$$

$$[\psi_\sigma(\vec{r}), \psi_{\sigma'}(\vec{r}')]_+ = [\psi_\sigma^+(\vec{r}), \psi_{\sigma'}^+(\vec{r}')]_+ = 0$$

(2.1.3)

We chose [ , ]$_+$ instead of [ , ]$_-$, since Hamiltonian must be positive definite.

If we write the Lagrangian and Hamiltonian density as

$$L = -\psi_\sigma^+(\vec{r})(-\frac{\nabla^2}{2m} + V(\vec{r}))\psi_\sigma(\vec{r}) + i\frac{\partial \psi_\sigma^+}{\partial t}\psi_\sigma \quad (2.1.4)$$

$$H = \Pi_{\psi_\sigma^+} \frac{\partial \psi_\sigma^+}{\partial t} - L \quad (2.1.5)$$

Where

$$\Pi_{\psi_\sigma^+} = \frac{\partial L}{\partial(\partial \psi_\sigma^+ / \partial t)} \qquad (2.1.6)$$

The Hamiltonian will be

$$H = \sum_\sigma \int d^3 r (Hamiltonian - density) = \sum_\sigma \int d^3 r \psi_\sigma^+(\vec{r})(-\frac{\nabla^2}{2m} + V(\vec{r}))\psi_\sigma(\vec{r}) \qquad (2.1.7)$$

Therefore in terms of them, the Hamiltonian $H$ takes the form

$$H = \sum_\sigma \int d^3 r \psi_\sigma^+(\vec{r})(-\frac{1}{2m}\nabla^2 + V(\vec{r}))\psi_\sigma(\vec{r}) + \frac{e^2}{2}\sum_{\sigma\sigma'}\int d^3 r d^3 r' \psi_\sigma^+(\vec{r})\psi_\sigma(\vec{r})\frac{1}{|\vec{r}-\vec{r'}|}\psi_{\sigma'}^+(\vec{r'})\psi_{\sigma'}(\vec{r'})$$

$$(2.1.8)$$

In most cases one is trying to find approximate eigenstates of $H$ within a given set of L basis functions $f_i(\vec{r})$. If $a_{j\sigma}^+, a_{j\sigma}$ are electron creation and annihilation operators, an approximation for $\psi_\sigma(\vec{r})$ can be written as:

$$\psi_\sigma(\vec{r}) = \sum_{i=1}^L a_{i\sigma} f_i(\vec{r}) \qquad (2.1.9)$$

The functions $f_i(\vec{r})$ are generally not orthogonal to each other. Their overlap matrix is defined by:

$$S_{ij} = \int d^3 r f_i^*(\vec{r}) f_j(\vec{r}) \qquad (2.1.10)$$

From (2.1.4) and (2.1.6), it follows that

$$\left[a_{i\sigma}^+, a_{j\sigma'}\right]_+ = S_{ji}^{-1}\delta_{\sigma\sigma'}$$

$$\left[a_{i\sigma}^+, a_{j\sigma'}^+\right]_+ = \left[a_{i\sigma}, a_{j\sigma'}\right]_+ = 0 \qquad (2.1.11)$$

Within the space spanned by the function $f_i(\vec{r})$, the Hamiltonian (2.1.4) becomes

$$H = \sum_{ij\sigma} t_{ij} a_{i\sigma}^+ a_{j\sigma} + \frac{1}{2}\sum_{\substack{ijkl \\ \sigma\sigma'}} V_{ijkl} a_{i\sigma}^+ a_{k\sigma'}^+ a_{l\sigma'} a_{j\sigma} \qquad (2.1.12)$$

Where the matrices $t_{ij}$ and $V_{ijkl}$ are

$$t_{ij} = \int d^3 r f_i^*(\vec{r})(-\frac{1}{2m}\nabla^2 + V(\vec{r}))f_j(\vec{r})$$

$$(2.1.13)$$

$$V_{ijkl} = e^2 \int d^3 r d^3 r' f_i^*(\vec{r}) f_j(\vec{r}) \frac{1}{|\vec{r}-\vec{r'}|} f_k^*(\vec{r'}) f_l(\vec{r'})$$

In the following discussion, a Hamiltonian of the form (2.1.12) will be used frequently.

Finding the eigenfunctions and eigenvalues of an electronic system described by the Hamiltonian (2.1.12) is impossible without drastic simplifying assumptions. Hartree [6], Fock [7], and Slater [8] treated the electrons as being independent of each other and introduced the idea of the **Self-Consistent Field (SCF)**. The latter is the interaction fields an electron experiences when we take a spatial average over the positions of all the other electrons. Within the independent-electron approximation, we can assume the wave function $\varphi_1$ can be exist at $r_1, r_2,\ldots r_N$, the wave function $\varphi_2$ can be exist at $r_1, r_2,\ldots r_N$ and so on i.e we have a charge distribution. In this condition the ground state wave function is of the form

$$\Phi(\vec{r}_1\sigma_1,\ldots,\vec{r}_N\sigma_N) = \frac{1}{\sqrt{N!}} \begin{vmatrix} \varphi_1(\vec{r}_1\sigma_1) & \ldots & \varphi_N(\vec{r}_N\sigma_N) \\ \ldots & & \ldots \\ \ldots & & \ldots \\ \varphi_1(\vec{r}_N\sigma_N) & \ldots & \varphi_N(\vec{r}_N\sigma_N) \end{vmatrix} \qquad (2.1.14)$$

Without loss of generality, the functions $\varphi_\nu(\vec{r}\sigma)$ can be assumed to be orthogonal to each other.

On the other hand, we want not only the expectation value for the ground-state energy, $E_0 = \prec \Phi|H|\Phi \succ$, is minimized by variation of $\varphi_\mu(\vec{r}\sigma)$ and $\varphi_\mu^*(\vec{r}\sigma)$ but also the constraint $\prec \varphi_\mu|\varphi_\mu \succ = 1$ is satisfied. These conditions are taken into account by introducing Lagrange parameters $\varepsilon_\mu$ when doing the variation. The requirement is therefore

$$\delta(E_0 - \sum_\mu \varepsilon_\mu \prec \varphi_\mu|\varphi_\mu \succ) = 0 \qquad (2.1.15)$$

But

$$E_0 = \sum_{\nu=1}^N \prec \varphi_\nu|h|\varphi_\nu \succ + \frac{1}{2}\sum_{\mu\nu}^N [\prec \varphi_\mu\varphi_\nu|v|\varphi_\mu\varphi_\nu \succ - \prec \varphi_\mu\varphi_\nu|v|\varphi_\nu\varphi_\mu \succ] \qquad (2.1.16)$$

Where

$$\prec \varphi_\nu|h|\varphi_\nu \succ = \int d^3r\, \chi_\mu^*(\vec{r})h(\vec{r})\chi_\nu(\vec{r}) \qquad (2.1.17)$$

$$\prec \varphi_\mu\varphi_\tau|v|\varphi_\nu\varphi_\rho \succ = \delta_{\sigma_\mu\sigma_\nu}\delta_{\sigma_\tau\sigma_\rho} \int d^3r\, d^3r'\, \chi_\mu^*(\vec{r})\chi_\nu(\vec{r})v(\vec{r},\vec{r}')\chi_\tau^*(\vec{r}')\chi_\rho(\vec{r}') \qquad (2.1.18)$$

and $\varphi_\mu(\vec{r}) = \chi_\mu(\vec{r})\sigma$ are a product of a spatial orbital $\chi_\mu(\vec{r})$ and a two-component spinor σ. The latter equal equals $\alpha = \begin{pmatrix}1\\0\end{pmatrix}$ for spin-up and $\beta = \begin{pmatrix}0\\1\end{pmatrix}$ for spin-down electrons with respect to a preferred axis.

By substituting (2.1.16) in (2.1.15), it can be shown

$$F|\varphi_\mu \succ = \varepsilon_\mu|\varphi_\mu \succ \qquad (2.1.19)$$

Where F is the Fock operator with matrix element

$$f_{\nu\mu} = \prec \varphi_\nu |h| \varphi_\mu \succ + \sum_\tau^N (\prec \varphi_\nu \varphi_\tau |v| \varphi_\mu \varphi_\tau \succ - \prec \varphi_\nu \varphi_\tau |v| \varphi_\tau \varphi_\mu \succ) \quad (2.1.20)$$

Equation (2.1.19) constitutes the well-known **Hartree-Fock (HF) equations**. The eigenvalues $\varepsilon_\mu$ are obtained from the diagonal form of the Fock matrix and the total energy E is not simply the sum over $\varepsilon_\mu$ but, instead, given by

$$E_0 = \sum_\mu^N \varepsilon_\mu - \frac{1}{2} \sum_{\mu\nu}^N (\prec \varphi_\mu \varphi_\nu |v| \varphi_\mu \varphi_\nu \succ - \prec \varphi_\mu \varphi_\nu |v| \varphi_\nu \varphi_\mu \succ) \quad (2.1.21)$$

Solving the HF equation exactly is not possible except in trivial cases such as that of the homogeneous electron gas. For inhomogeneous systems calculations are done with a set of basis functions $f_i(\vec{r})$, in which case the Hamiltonian takes the form of (2.1.12). In order to derive the Fock operator here, the determinant (2.1.14) is written in second quantized form as

$$|\Phi \succ = \prod_{\mu\sigma} c^+_{\mu\sigma} |0 \succ \quad (2.1.22)$$

Where $c^+_{\mu\sigma}$ is the creation operator for an electron in spin orbital $\varphi_\mu(\vec{r}\sigma)$ and $|0\succ$ the vacuum state.

Because the $\varphi_\mu(\vec{r}\sigma)$ can be assumed to be orthogonal, the $c^+_{\mu\sigma}$ satisfy the usual fermionic anticommutation relations

$$[c^+_{\mu\nu'}, c_{\nu\sigma'}]_+ = \delta_{\mu\nu} \delta_{\sigma\sigma'}$$
$$\quad (2.1.23)$$
$$[c^+_{\mu\sigma}, c^+_{\nu\sigma'}] = [c_{\mu\sigma}, c_{\nu\sigma'}]_+ = 0$$

The wave function (2.2.21) is denoted by $|\Phi_{SCF} \succ$ in the text below, the corresponding expectation value of the Hamiltonian (2.1.12) is given by

$$E_0 = \sum_{ij}^L \sum_\sigma t_{ij} \prec \Phi_{SCF} |a^+_{i\sigma} a_{j\sigma}| \Phi_{SCF} \succ + \frac{1}{2} \sum_{ijkl} \sum_{\sigma\sigma'} V_{ijkl} \prec \Phi_{SCF} |a^+_{i\sigma} a^+_{k\sigma'} a_{l\sigma'} a_{j\sigma}| \Phi_{SCF} \succ \quad (2.1.24)$$

The occupied orbital of the ground state $|\Phi_{SCF}\succ$ are expanded in terms of the basis functions $f_i(\vec{r})$ as

$$c^+_{\mu\sigma} = \sum_{n=1}^L d_{\mu n} \hat{a}^+_{n\sigma} \quad (2.1.25)$$

and the bond-order matrix, $P_{ij}$ is defined as

$$P_{ij} = \sum_\sigma \prec a^+_{i\sigma} a_{j\sigma} \succ = 2 \sum_\nu^{occ} d^*_{\nu i} d_{\nu j} \quad (2.1.26)$$

Now if we repeat the condition (2.1.15), we will find

$$\sum_{j=1}^L (f_{ij} - \varepsilon_\mu S_{ij}) d_{ij} = 0 \quad (2.1.27)$$

Where the $f_{ij}$ are given by

$$f_{ij} = t_{ij} + \sum_{kl}(V_{ijkl} - \frac{1}{2}V_{ilkj})P_{kl} \quad (2.1.28)$$

These are the **Self-Consistent Field (SCF) equations** for a finite basis set [9].

We can satisfy (2.1.26) if we introduce the Fock operator

$$F = \sum_{ij\sigma} f_{ij}(a_{i\sigma}^+ a_{j\sigma} - \prec a_{i\sigma}^+ a_{j\sigma} \succ) \quad (2.1.29)$$

and require that the following self-consistent relation hold

$$Fc_{\mu\sigma}|\Phi_{SCF}\succ = -\varepsilon_\mu c_{\mu\sigma}|\Phi_{SCF}\succ$$

$$\quad (2.1.30)$$

$$Fc_{i\sigma}^+|\Phi_{SCF}\succ = \varepsilon_i c_{i\sigma}^+|\Phi_{SCF}\succ$$

depending on whether the coefficient for occupied $(d_{\mu n})$ or unoccupied $(d_{in})$ orbital are considered. The corresponding spin orbital $\varphi_\mu(\vec{r}\sigma)$ and $\varphi_i(\vec{r}\sigma)$ are the canonical molecular orbital (MOs). It can be checked that $F|\Phi_{SCF}\succ = 0$.

The solution of the SCF equation must fulfill symmetry requirements which restrict their form. For example, in addition to being a solution of (2.1.26), a molecular orbital must be an eigenfunction of the different symmetry operations with which the Hamiltonian commutes, i.e., it must transform according to an irreducible representation of the point group of the molecule.

In addition to the symmetry requirements, there are also equivalence restrictions. An equivalence restriction is, for example, that the orbital function of $4\sigma_\uparrow$ is the same as that of $4\sigma_\downarrow$. The equivalence restrictions are consequences of conservation laws.

**Unrestricted Hartree-Fock wave functions** break both symmetry and equivalence requirements. With these wave functions one often obtain lower energies than when working with restricted Hartree-Fock functions. The reason is that breaking a symmetry implies that electronic correlations are partially taken into account.

### b. Density Functional Theory

The density functional approach expresses ground-state properties- such as total energy, equilibrium positions and moments- in terms of the electronic density $\rho(\vec{r})$ or spin density $\rho_\sigma(\vec{r})$ and provides a scheme for calculating them. The method avoids the problem of calculating the ground-state wave function.

**Thomas-Fermi Method**

Consider a charge distribution of electron, $\rho(\vec{r})$. If $V(\vec{r})$ is a given external electrostatic potential, the energy of an electronic density $\rho(\vec{r})$ in this potential will be equal to $\int d^3r V(\vec{r})\rho(\vec{r})$. Also for a classical charge distribution the repulsion energy is equal to $\frac{e^2}{2}\int d^3r d^3r' \frac{\rho(r)\rho(\vec{r'})}{|\vec{r}-\vec{r'}|}$ and the kinetic energy density of homogenous system is equal to $\frac{3}{10m}(3\Pi^2)^{2/3}\int d^3r\rho^{5/3}(\vec{r})$. The crucial assumption of the Thomas-Fermi method is the form of the ground-state energy functional $E_{TF}[\rho,V]$, which one writes as [10, 11]

$$E_{TF}[\rho,V] = \int d^3r V(\vec{r})\rho(\vec{r}) + \frac{e^2}{2}\int d^3r d^3r' \frac{\rho(\vec{r})\rho(\vec{r}')}{|\vec{r}-\vec{r}'|} + \frac{3}{10m}(3\Pi^2)^{2/3}\int d^3r\rho^{5/3}(\vec{r}) \quad (2.2.1.1)$$

Now we have two conditions

$$\delta E_{TF}[\rho,V] = 0 \text{ i.e. } E_{TF}[\rho,V] \text{ is minimum}$$
$$\delta\int d^3r\rho(\vec{r}) = 0 \text{ i.e. total electron number remains constant} \quad (2.2.1.2)$$

By using Lagrange parameter μ, the requirement is therefore

$$\delta(E_{TF}[\rho,V] - \mu\int d^3r\rho(\vec{r})) = 0 \quad (2.2.1.3)$$

By substituting (2.2.1) in (2.2.3), one finds

$$V(\vec{r}) + e^2\int d^3r' \frac{\rho(\vec{r}')}{|\vec{r}-\vec{r}'|} + \frac{1}{2m}(3\Pi^2)^{2/3} - \mu = 0 \quad (2.2.1.4)$$

By introducing the effective potential, $V_c(\vec{r})$, as

$$V_c(\vec{r}) = e^2\int d^3r' \frac{\rho(\vec{r}')}{|\vec{r}-\vec{r}'|}$$
$$\nabla^2 V_c(\vec{r}) = -4\Pi e^2\rho(\vec{r}) \quad (2.2.1.5)$$

One can writes

$$\rho(\vec{r}) = \frac{(2m)^{3/2}}{3\Pi^2}[\mu - V(\vec{r}) - V_c(\vec{r})]^{3/2} \quad (2.2.1.6)$$

Equations (2.2.1.5) and (2.2.1.6) yield a differential equation for $V_c(\vec{r})$. Here we neglected the exchange and correlation effects.

**Hohenberg-Kohn-Sham Theory**

There are two important theorems, called Hohenberg and Kohn theorems [12]

a)- The ground-state energy, E, of a many-electron system in the presence of an external potential $V(\vec{r})$ is a functional of the electronic density $\rho(\vec{r})$, and one can writes

$$E[\rho,V] = \int d^3r V(\vec{r})\rho(\vec{r}) + F(\rho) \quad (2.2.2.1)$$

Where $F(\rho)$ is an unknown, but universal functional of density $\rho(\vec{r})$ only, and does not depend on $V(\vec{r})$.

b)- The $E[\rho,V]$ is minimized by the ground state density.

Now if one assumes

$$F[\rho] = \frac{e^2}{2}\int d^3r d^3r' \frac{\rho(\vec{r})\rho(\vec{r}')}{|\vec{r}-\vec{r}'|} + T_0[\rho] + E_{xc}[\rho] \quad (2.2.2.2)$$

where the first term is repulsive energy, the second term is kinetic energy and the third term is exchange and correlation energy.

By using (2.2.1.2) one can find

$$\int d^3r \delta\rho(\vec{r})\left\{V(\vec{r}) + e^2\int d^3r' \frac{\rho(\vec{r}')}{|\vec{r}-\vec{r}'|} + \frac{\delta T_0[\rho]}{\delta\rho(\vec{r})} + \frac{\delta E_{xc}[\rho]}{\delta\rho(\vec{r})}\right\} = 0 \quad (2.2.2.3)$$

Now if one define the effective potential as

$$V_{eff}(\vec{r}) = V(\vec{r}) + e^2\int d^3r' \frac{\rho(\vec{r}')}{|\vec{r}-\vec{r}'|} + V_{xc}(\vec{r})$$

$$V_{xc}(\vec{r}) = \frac{\delta E_{xc}[\rho]}{\delta\rho(\vec{r})}$$

(2.2.2.4)

instead of solving many body equation, he/she can solve the following single electron Schrödinger equation

$$[-\frac{1}{2m}\nabla^2 + V_{eff}(\vec{r})]\chi_\mu(\vec{r}) = \varepsilon_\mu \chi_\mu(\vec{r})$$

$$\rho(\vec{r}) = 2\sum_\mu^{N/2}|\chi_\mu(\vec{r})|^2$$

(2.2.2.5)

The sum is over the eigenfuctions with the lowest eigenvalues. This equation is called **Kohn-Sham equation**.

A comment is in order on the physical significance of the eigenvalues $\varepsilon_\mu$ of (2.2.2.5). We have here

$$2\sum_\mu^{N/2}\varepsilon_\mu = T_0[\rho] + \int d^3r V_{eff}(\vec{r})\rho(\vec{r}) \quad (2.2.2.6)$$

From (2.2.2.1) and (2.2.2.2) it follows that the total energy is given by

$$E[\rho,V] = 2\sum_\mu^{N/2}\varepsilon_\mu - \frac{e^2}{2}\int d^3r d^3r' \frac{\rho(\vec{r})\rho(\vec{r}')}{|\vec{r}-\vec{r}'|} + E_{xc}[\rho] - \int d^3r v_{xc}(\vec{r})\rho(\vec{r}) \quad (2.2.2.7)$$

This relation can be compared with the corresponding one (2.1.21) for independent electrons. The real eigenvalues $\varepsilon_\mu$ do not describe electronic excitation energies, generally understood to be complex quantities due to finite lifetimes of excitations. However, it turns out that for infinite systems with extended states the energy of the highest occupied level is equal to the chemical potential.

## Local Density Approximation (LDA)

The Local Density Approximation (LDA) consists of replacing exchange correlation energy $E_{xc}[\rho]$ by

$$E_{xc}[\rho] = \int d^3r \rho(\vec{r}) \varepsilon_{xc}(\rho(\vec{r})) \quad (2.2.3.1)$$

where $\varepsilon_{xc}(\rho(\vec{r}))$ is the exchange and correlation energy per electron of a homogenous electron gas of density ρ, considered to be known. Therefore

$$V_{xc}(\vec{r}) = \frac{d[\rho(\vec{r})\varepsilon_{xc}(\rho(\vec{r}))]}{d\rho(\vec{r})} \quad (2.2.3.2)$$

and $V_{eff}(\vec{r})$ depends only on $\rho(\vec{r})$ and the Kohn-Sham equation take simpler form than Hartree-Fock equation.

## Local Spin Density Approximation (LSDA)

We can add the effect of spin to LDA, if we write $\rho = \rho_\uparrow + \rho_\downarrow$, where ↑ stands for spin up and ↓ stands for spin down. The exchange correlation energy is $\varepsilon_{xc}(\rho\uparrow, \rho\downarrow)$ and the (2X2) matrix equation, reduces to two coupled equations written as

$$(-\frac{1}{2m}\nabla^2 - \mu_B \vec{\sigma}.H(\vec{r}) + V_\sigma^{eff}(\vec{r}))\phi_{\mu\sigma}(\vec{r}) = \varepsilon_{\mu\sigma}\phi_{\mu\sigma}(\vec{r}) \quad (2.2.4.1)$$

The Zeeman term has been included, and $H(\vec{r})$ is an applied uniform magnetic field. The spin dependent effective single particle potential is given by

$$V_\sigma^{eff}(\vec{r}) = V(\vec{r}) + e^2 \int d^3r' \frac{\rho(\vec{r}')}{|\vec{r}-\vec{r}'|} + V_\sigma^{xc}(\vec{r}) \quad (2.2.4.2)$$

Where

$$V_\sigma^{xc}(\vec{r}) = \frac{d}{d\rho_\sigma(\vec{r})}[\rho_\uparrow(\vec{r}) + \rho_\downarrow(\vec{r})]\varepsilon_{xc}(\rho_\uparrow(\vec{r}), \rho_\downarrow(\vec{r})) \quad (2.2.4.3)$$

The spin densities are obtained from the functions $\phi_{\mu\sigma}(\vec{r})$ through

$$\rho_\sigma(\vec{r}) = \sum_\mu^{occ} |\phi_{\mu\sigma}(\vec{r})|^2 \quad (2.2.4.4)$$

The sum is over all occupied orbitals with spin σ.

### c. Electron correlations in semiconductors

Electron correlations are very important in the study of semiconductors. Correlation effects appear even more dramatic when excited states are considered. The energy gaps in semiconductors calculated within the SCF or HF approximation come out much too large and this is easy to understand in terms of the correlation picture. When an electron is excited from a valance into a conduction band, the added particle polarizes the bonds in its neighborhood because the system is locally no longer charge neutral. The polarized bonds form a polarization cloud which moves with the extra electron (hole) and together they form a quasiparticle. It costs much less energy to create an electron-hole pair with a polarization cloud than without. The generation of such polarization cloud is a correlation effect, which is not taken into account in the independent electron approximation.

In LDA to density functional theory, one can not distinguish the correlation holes around the added and the remaining electrons. The density dependent exchange correlation potential remains unchanged when an extra electron is added to the infinite system. The so-called energy gap problem finds its origin here. One can circumvent it by applying the local ansatz to the computation of energy bands in semiconductors.

### 2.3.1 Ground State Correlation

Correlation energy calculation for covalent semiconductors become very simple, and in fact can be performed analytically, when the bond-orbital approximation (BOA) is made. In the discussion below, a diamond-like structure is considered. Our starting point is a Hamiltonian of the form

$$H = \sum_{ij\sigma} t_{ij} a_{i\sigma}^+ a_{j\sigma} + \frac{1}{2} \sum_{\substack{ijkl \\ \sigma\sigma'}} V_{ijkl} a_{i\sigma}^+ a_{k\sigma'}^+ a_{l\sigma'} a_{j\sigma} \quad (2.3.1.1)$$

where

$$V_{ijkl} = \int d^3r d^3r' f_i^*(\vec{r}) f_j(\vec{r}) \frac{e^2}{|\vec{r}-\vec{r'}|} f_k^*(\vec{r'}) f_l(\vec{r'}) \quad (2.3.1.2)$$

and $f_j(\vec{r})$ are the set of L basis function which within them, the approximate eigenstate of H is found. The atomic $sp^3$ hybrid functions take place of the basis function $f_j(\vec{r})$. The $t_{ij}$ are parameters which can be determined from more sophisticated band-structure calculations.

The $a_{i\sigma}^+$ are creation operators for electrons in orthognalized, tetrahedral atomic hybrids. Hence, they fulfill the relations

$$[a_{i\sigma}, a_{j\sigma'}^+] = \delta_{ij}\delta_{\sigma\sigma'} \quad (2.3.1.2)$$

In the BOA all calculations are carried out in terms of bonding and anti-bonding wavefunctions with their corresponding creation operators

$$B_{I\sigma}^+ = \frac{1}{\sqrt{2}}(a_{I1\sigma}^+ + a_{I2\sigma}^+)$$

$$A_{I\sigma}^+ = \frac{1}{\sqrt{2}}(a_{I1\sigma}^+ - a_{I2\sigma}^+) \quad (2.3.1.3)$$

Here I is a bond index. The two hybrids forming bond I are indexed by the subscripts 1 and 2, respectively. The SCF ground state is assumed to be the form

$$|\Phi_{BOA}\succ = \prod_{I\sigma} B_{I\sigma}^+ |0\succ \quad (2.3.1.4)$$

i.e. the one particle density matrix $p_{ij}^{\sigma\sigma'}$ is simply

$$p_{ij}^{\sigma\sigma'} = \prec a_{i\sigma}^+ a_{j\sigma'} \succ = \begin{cases} \delta_{\sigma\sigma'}/2 \to for\, i.j \in I \\ 0 \to Otherwise. \end{cases} \quad (2.3.1.5)$$

There are two electrons in each bond.

The Hilbert space within which we calculate the correlated ground state $|\psi_0\succ$ is spanned by the state $|\Phi_{BOA}\succ$, $A_{I\sigma}^+ B_{I\sigma}|\Phi_{BOA}\succ$ and $A_{I\sigma}^+ A_{J\sigma'}^+ B_{J\sigma'} B_{I\sigma}|\Phi_{BOA}\succ$.

For correlated ground state wavefunction, one can make the ansatz

$$|\psi_0\succ = e^S |\Phi_{BOA}\succ$$

$$|\psi_0\succ = (1 + \sum_{i\mu} \alpha_\mu^i c_i^+ c_\mu + \sum_{\substack{i\prec j \\ \mu\prec\theta}} \alpha_{\mu\nu}^{ij} c_i^+ c_j^+ c_\nu c_\mu + ...)|\Phi_{BOA}\succ \quad (2.3.1.6)$$

$$|\psi_0\succ = (1 + \sum_{i\mu} \alpha_\mu^i \omega_\mu^i + \sum_{\substack{i\prec j \\ \mu\prec\nu}} \alpha_{\mu\nu}^{ij} \omega_{\mu\nu}^{ij} + ...)|\Phi_{BOA}\succ$$

Here $\omega_\mu^i = c_i^+ c_\mu$, $\omega_{\mu\nu}^{ij} = c_i^+ c_j^+ c_\nu c_\mu$, and so on. The $c_i, c_i^+$ are defined in (2.1.25). We obtain the $\alpha_\mu^i, \alpha_{\mu\nu}^{ij}, etc.$ diagonalizing H, within a Hilbert space of a given dimension. Alternatively, one may consider the coefficient as variational parameters fixed by minimizing the energy $E = \prec \psi_0|H|\psi_0\succ / \prec \psi_0|\psi_0\succ$.

The ground state energy is equal to

$$E_0 = \prec H \succ + \sum_{i\mu} \prec H\omega_\mu^i \succ \alpha_\mu^i + \sum_{\substack{i\prec j \\ \mu\prec\nu}} \prec H\omega_{\mu\nu}^{ij} \succ \alpha_{\mu\nu}^{ij}$$

Or

$$E_0 = \prec H \succ + \sum_{\mu\prec\nu} E_{\mu\nu},$$

$$E_{\mu\nu} = \sum_{i\prec j} \prec H\omega_{\mu\nu}^{ij} \succ \alpha_{\mu\nu}^{ij} \quad (2.3.1.7)$$

The electrons are annihilated and created in localized molecular (as well as local) orbitals, therefore $\omega_\mu^i = c_i^+ c_\mu$, $\omega_{\mu\nu}^{ij} = c_i^+ c_j^+ c_\nu c_\mu$, etc. and $c_i, c_i^+$ refer to localized molecular oebitals.

### 2.3.2 Excited States

Electron correlations in semiconductors with electrons in excited states show features which are absent in the ground state, a fact intimately connected with the so-called energy-gap problem. While calculations done within LDA to the density functional method always seem to give too small energy gaps when applied to semiconductors, the form of different bands agrees quite well with experiment. Our starting point is again the Hamiltonian (2.3.1.1) with a SCF part

$$H_{SCF} = \sum_{ij\sigma} f_{ij}(a^+_{i\sigma}a_{j\sigma} - \prec a^+_{i\sigma}a_{j\sigma} \succ) + E_{SCF} \qquad (2.3.2.1)$$

It will turn out to be useful to separate the Fock matrix $f_{ij}$ into two parts

$$f_{ij} = f^H_{ij} + f^x_{ij} \qquad (2.3.2.2)$$

Where

$$f^H_{ij} = t_{ij} + \sum_{kl\sigma'} V_{ijkl} \prec a^+_{k\sigma'}a_{l\sigma'} \succ \qquad (2.3.2.3)$$

is the Hartree part and

$$f^x_{ij} = -\frac{1}{2}\sum_{kl\sigma'} V_{ijkl} \prec a^+_{k\sigma'}a_{l\sigma'} \succ \qquad (2.3.2.4)$$

is the exchange part.
By using (2.3.1.2), the notation

$$\left.\begin{array}{l} U = V_{iiii} \\ J_1 = V_{iijj} \\ K_1 = V_{ijji} \end{array}\right\} i \neq j, i, j \in I \qquad (2.3.2.5)$$

is introduced. Here I is a bond index.
The SCF eigenstate of the (N+1)-electron system is written in the form

$$|\Phi_{kc\sigma} \succ = c^+_{kc\sigma}|\Phi_{SCF} \succ \qquad (2.3.2.6)$$

where $c^+_{kc\sigma}$ creates the extra electron in conduction band c with momentum **k** and spin σ. The corresponding SCF energy is

$$[H_{SCF}, c^+_{kc\sigma}]_- = \varepsilon^{SCF}_{c\sigma}(k)c^+_{kc\sigma} \qquad (2.3.2.7)$$

We apply the BOA in order to evaluate the effect of exchange on the energy bands. Now the exchange can be evaluated analytically because the one particle density matrix is of the simple form of (2.3.1.5). The SCF conduction and valanced bands are obtained from

$$\begin{aligned} [H_{SCF}, c^+_{kc\sigma}]_- &= \varepsilon^{SCF}_{c}(k)c^+_{kc\sigma} \\ [H_{SCF}, c_{kv\sigma}]_- &= -\varepsilon^{SCF}_{v}(k)c_{kv\sigma} \end{aligned} \qquad (2.3.2.8)$$

One can writes

$$\varepsilon_c^{SCF}(k) = (f_{ii}^H - \frac{U}{2} - \frac{3}{2}V_{1331} - ...) - (f_{12}^H - \frac{J_1}{2} - \frac{1}{2}K_1 - ...) + \varepsilon_c^{\sim SCF}(k) \quad , c=1,...4$$

(2.3.2.9)

$$\varepsilon_\nu^{SCF}(k) = (f_{ii}^H - \frac{U}{2} - \frac{3}{2}V_{1331} - ...) - (f_{12}^H - \frac{J_1}{2} - \frac{1}{2}K_1 - ...) + \varepsilon_\nu^{\sim SCF}(k) \quad , c=1,...4$$

The $\varepsilon^{\sim SCF}(k)$ differ fro the $\varepsilon^{\sim H}(k)$ by the inclusion of exchange contributions. The energies U and $J_1$, the dominant interaction matrix elements, are defined by (2.3.2.5).

The effect of correlation can be calculated by making use of the projection method. Let $|\psi_0^N \succ$ denote the ground state of an elemental semiconductor with energy $E_0^{(N)}$. When an electron with momentum **k** and spin σ is put into a conduction band c, the energy of the exact (N+1) electron eigenstate is $E_0^{(N+1)}(k,c)$. The excitation energy is defined by

$$\varepsilon_c(k) = E_0^{(N+1)}(k,c) - E_0^{(N)} \quad (2.3.2.10)$$

and is contained in the one particle correlation function

$$R_\sigma(k,c,\tau) = \prec \psi_0 | c_{kc\sigma} e^{-\tau[H-E_0^{(N)}]} c_{kc\sigma}^+ | \psi_0^{(N)} \succ \quad (2.3.2.11)$$

Within the quasiparticle approximation this expression reduces to

$$R_\sigma(k,c,\tau) = \prec \psi_0^{(N)} | c_{kc\sigma} c_{kc\sigma}^+ | \psi_0^{(N)} \succ e^{-\tau\varepsilon_c(k)} \quad (2.3.2.12)$$

where $\varepsilon_c(k)$ is the quasiparticle energy. It shows up as a pole the Laplace transform $R_\sigma(k,c,z)$, i.e.

$$R_\sigma(k,c,z) = (c_{kc\sigma}\Omega | \frac{1}{z-(L_{SCF}+H_{res})} c_{kc\sigma}^+\Omega) \quad (2.3.2.13)$$

The operator $\Omega$ describes the exact ground state $|\psi_0^N \succ$. The $L_{SCF}$ is Liouville operator belonging to $H_{SCF}$ and defined through $L_0 A = [H_{SCF}, A]_-$ where A is arbitrary operator. We assumed $H = H_{SCF} + H_{res}$, the eigenstates and eigenvalues of $H_{SCF}$ are known and the effect of $H_{res}$ on the ground state energy is relatively small. The excitation energies are given by the poles of $R_\sigma(k,c,z)$ [13].

## 2.4 Relativistic Effective Core Potential (RECP)

One of the most fundamental assumptions in chemistry is that low lying core electrons are relatively inert, and are not perturbed by a molecular environment. This assumption is supported by the chemical similarity of elements in the same column of the periodic table. Most of the important chemical properties of atoms and molecules are determined by the interaction of their valence electrons with the valence electrons of other atomic or molecular species. For molecules containing heavy atoms, large computational savings may be realized if a particular mathematical form for the atomic orbitals is assumed, *a*

*priori*, so that the total number of parameters which must be optimized in, say, the computation of a Hartree-Fock wavefunction may be reduced. This approximation on its own is known as the frozen core approximation. A more extensive savings may be achieved if, instead of simply assuming a particular form for the core orbitals, all terms describing the interaction of the electrons in these orbitals with each other and those outside the core region are simply replaced by a scalar ``effective potential''. This provides the additional benefit of reducing the basis set size requirements, since no basis functions are now needed to explicitly describe the core orbitals. This has the effect of drastically reducing the number of two-electron quantities, such as electron repulsion integrals, which must be considered. In correlated electronic structure calculations, the molecular orbitals which correspond to atomic core orbitals are often excluded from the active orbital space. Even when such orbitals are included in the active space, the resultant contributions are typically small for most chemical properties. The use of effective core orbitals in these methods ideally should not significantly affect the quality of the resulting molecular property predictions. Because correlated methods scale higher than Hartree-Fock with respect to basis set size, the decreased basis set size resulting from the use of effective potentials will introduce in an even more dramatic computational savings over all-electron methods.

By using (2.1.8), the Fock operator is defined as

$$\hat{F} = -\frac{\nabla_r^2}{2} - \frac{Z}{r} + \frac{l(l+1)}{2r^2} + \sum_{j=1}^{n}(\hat{J}_i - \hat{K}_j) \quad (2.4.1)$$

where the action of the operator $\hat{J}_j$ and $\hat{K}_j$ is defined by

$$\hat{J}_j \phi_i(\vec{r_1}) = (\int dr_2 \phi_j^*(\vec{r_2}) \frac{1}{r_{12}} \phi_j(\vec{r_2}))\phi_i(\vec{r_1})$$
$$\hat{K}_j \phi_i(\vec{r_1}) = (\int dr_2 \phi_j^*(\vec{r_2}) \frac{1}{r_{12}} \phi_i(\vec{r_2}))\phi_j(\vec{r_1}) \quad (2.4.2)$$

The $\phi_i$ are the single particle eigenfunctions of the Fock operator.
If we now divide the orbitals into a group of $N_c$ core orbitals and $V_v$ valence orbitals, we may re-express the fock operator as

$$\hat{F} = -\frac{\nabla_r^2}{2} - \frac{Z}{r} + \frac{l(l+1)}{2r^2} + \hat{V}^{core} + \hat{V}^{val} \quad (2.4.3)$$

Where $\hat{V}^{core} = \sum_{a}^{N_c}(\hat{J}_a - \hat{K}_a)$ and $\hat{V}^{val} = \sum_{i}^{N_v}(\hat{J}_i - \hat{K}_i)$.

For the molecular case, one can expand the Fock operator, if he/she replaces $\hat{V}^{core}$ by an effective core potential (ECP), $\hat{V}^{eff}$, a local potential, and the nuclear charge, Z, by $Z_{eff} = Z - N_c$, where $N_c$ is the number of core electrons. The Fock equation will take the form

$$(-\frac{\nabla_r^2}{2} - \frac{Z_{eff}}{r} + \frac{l(l+1)}{2r^2} + \hat{V}^{eff} + \hat{V}^{val})\phi_i = \varepsilon_l \phi_i \quad (2.4.4)$$

The form of this effective potential is typically derived from numerical Hartree-Fock atomic solution, in accordance with the frozen core approximation.

If $\hat{l}$ is angular momentum operator of a valance atomic orbital, and we replace $\phi_i$ by approximate pseudo-orbitals, $\chi_i$, which are nodeless, we can rewrite (2.2.4) as

$$(-\frac{\nabla_r^2}{2} - \frac{Z}{r} + \frac{l(l+1)}{2r^2} + V_I^{eff} + V_{val}^I)\chi_i^I = \varepsilon_i^I \chi_i^I \qquad (2.4.5)$$

By solving (2.4.5) for $V_I^{eff}$ one find

$$V_I^{eff} = \varepsilon_i^I + \frac{Z_{eff}}{r} - \frac{l(l+1)}{2r^2} + \frac{(\frac{1}{2}\nabla_r^2 - V_{val}^I)\chi_i^I}{\chi_i^I} \qquad (2.4.6)$$

The total $V_{eff}$ for an atom, then, is written

$$V_{eff} = \sum_I V_I^{eff} \sum_{m=-I}^{I} |lm\succ\prec lm| \qquad (2.4.7)$$

So that the orthonormality of the spherical harmonics, $|lm\succ$, pairs each atomic wavefunction with proper $V_I^{eff}$ term.

Relativistic effective core potential (RECP) may be obtained in a manner directly analogous to the non-relativistic EDP's. The starting point for RECP's is the atomic Dirac-Coulomb-Fock equation. The valence solution, $\phi_i$, are four-component spinors, but the large radial component generally accounts for over 99% of the electronic density. Therefore the re-normalized large component are used as the starting points for the relativistic pseudo-orbitals, $\chi_i$. Once $\chi_i$ have been determined, the RECP's are obtained. The new form of (2.4.5) is

$$(-\frac{\nabla_r^2}{2} - \frac{Z_{eff}}{r} + \frac{l(l+1)}{2r^2} + \hat{V}_{I,j}^{eff} + \hat{V}_{val}^I)\chi_i^{I,j} = \varepsilon_i^{I,j} \chi_i^{I,j} \qquad (2.4.6)$$

where $\hat{j} = \hat{l} + \hat{s}$ is total angular momentum operator.

Since the RECP's are typically used within the framework of non-relativistic quantum chemistry, where single particle states are eigenfunction of $\hat{l}$ and $\hat{s}$ rather than $\hat{j}$, it is necessary to construct a $V^{eff}$ which is only $\hat{l}$ dependent. This may be accomplished by statistically averaging over all of the appreciate j-dependent RECP's which are associated with a particular I-value to give what are known as I-averaged RECP's (AREP).

$$V_I^{AREP} = (2l+1)^{-1}[lV_{l,j=l-\frac{1}{2}}^{REP} + (l+1)V_{l,j=l+\frac{1}{2}}^{REP}] \qquad (2.4.7)$$

The total $V^{AREP}$ may be given by

$$V^{AREP} = \sum_I V_I^{AREP} \sum_{m=-l}^{I} |lm\succ\prec lm| \qquad (2.4.8)$$

which is completely analogous to (2.4.7).

For atoms beyond the third row of the periodic table i.e for these very large nuclei, electrons near the nucleus are treated in an approximate way, via ECPs. This treatment includes some relativistic effects, which are important in these atoms.

## 2.5 Symmetry of Clusters

The space group of many of the simpler crystal structures are, referred to as point space groups [14]. Every element of a point space group may be written as the product of an element of a translation group with an element of a point group, every element of the latter being either a rotation or the product of a rotation and an inversion.

In clusters we deal with the point group only. The zinc blende structure has the point group $T_d^2$, and the face-centered cubic structure has the point space group $O_h^5$. The space group for the diamond structure, $O_h^7$, is not a point space group [14].

The lines and points of symmetry in the Brillouin zone of the zinc blende structure are shown in figure 1.

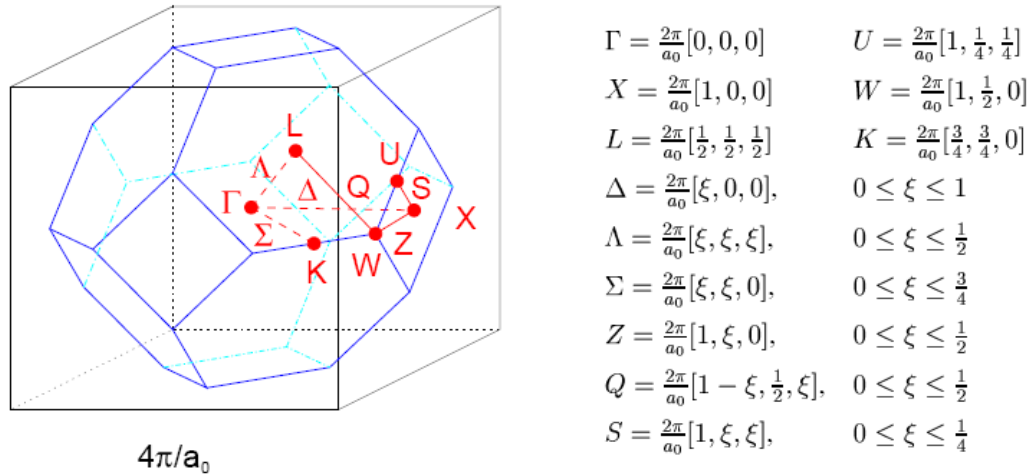

Fig.1. Lines and Points of symmetry in the Brillouin zone of the zinc blende structure

As an example, rotation by $90^0$ about the axis $\triangle$ followed by an inversion, followed by the time reversal symmetry operation (replacing $\vec{k}$ by $-\vec{k}$) will convert a wave function belonging to $\triangle_3$ into one belonging to $\triangle_4$, while leaving $\vec{k}$ and Hamiltonian invariant. Thus $\triangle_3$ and $\triangle_4$ stick together [15].

Also one can define the action integral as

$$S = \int L(\phi, \partial_\mu \phi, x_\mu) d^4 x \quad (2.5.1)$$

where $L(\phi, \partial_\mu \phi, x_\mu)$ is Largrangian density function.

If Under the transformation of space-time and internal space i.e

$$x^\mu \to x'^\mu$$
$$\phi(x) \to \phi'(x') \quad (2.5.2)$$

the action S is invariant, it can be shown that, there is a current density as follows

$$j_i^\nu = -L(X_i)_\mu^\nu + \frac{\partial L}{\partial U^A_{,\nu}}[U^A_{,\lambda}(X_i)_\mu^\lambda x^\mu - (X_i)^A_B U^B] \quad (2.5.3)$$

Where
$$x'^\mu = (X_i)_\mu^\nu x^\mu$$
$$U^A = (X_i)^A_B U^B \quad (2.5.4)$$

such that $\partial_\nu j^\nu = 0$. It is called Neother's theorem [16].
We can divide the $j_i^\nu$ to two parts as

$$j_i^\nu = [-L(X_i)_\mu^\nu x^\mu + \frac{\partial L}{\partial U^A_{,\nu}} U^A_{,\lambda}(X_i)_\mu^\lambda x^\mu] - [\frac{\partial L}{\partial U^A_{,\nu}}(X_i)^A_B U^B] \quad (2.5.5)$$

where first part is referred to space-time transformation and second part is referred to internal space transformation. By usage of the first part, one can find the conservation law of angular momentum or energy-momentum tensor and by usage of second part, he/she can find the conservation law of spin (i.e. internal angular momentum)[16]. Therefore the symmetry of cluster will appear itself in conservation laws and $\vec{k}$, and Hamiltonian invariant.

### 3. InSb Clusters

It is important to select an appropriate basis set and cluster to treat a particular problem. The clusters are terminated with hydrogen, with the X-H bonds oriented along the bulk bond direction. The cluster itself takes two main forms. The first is atom-centered surrounded by concentric shells of bulk atoms. This cluster therefore possesses a tetrahedral symmetry. For the III-V's, this means that there is an excess of either the group-III or group-V atom type, and consequentially the neutral cluster would have the incorrect number of bonding electrons. For all of the bonding orbital to be filled, the neutral charge state is modeled by charging the cluster by the difference in the number of, group-III and group-V atoms. For example, a Ga centered cluster $Ga_{19}As_{16}H_{36}$ would be charged 3e. The second type of cluster is bond centered, which avoids the charging problem encountered in the atom centered III-V cluster as these clusters contain the same number of group-III and group-V atoms (by symmetry). This type of cluster is referred to as stoichiometric, and possesses trigonal symmetry. One disadvantage of using stoichiometric cluster for III-V compound is that they possess an inherent dipole. Consequentially the central bond is longer than the six equivalent back bonds. However, for most III-V materials all bonds are reproduced to within 3% of the experimental value [17].

We considered $In_4Sb_4H_{18}$ cluster with zinc-blend structure of InSb and used Guassian 2003 Software[18] for doing simulation. The figure 2 shows the cluster. The cohesive energy and band gap of InSb were calculated by Unrestricted Hartee-Fock (UHF) and Density Functional Theory (DFT) method under UHF/STO-3G and B3LYP/STO-3G codes. The results were shown in table 1.

**Table 1. The cohesive energy and band gap of InSb, non-relativistic and relativistic cases**

|  | UHF | | DFT | |
|---|---|---|---|---|
|  | Cohesive energy (a.u) | Band gap (ev) | Cohesive energy(a.u) | Band gap (ev) |
| Non-relativistic case | 0.3165 | 4.3224 | 0.2952 | 0.9184 |
| Relativistic case | 0.1266 | 1.7049 | 0.1288 | 0.6509 |

It is noted that, the band gap was estimated by the difference in the energies of the highest occupied and lowest unoccupied Khon-Sham eigenvalues. For doing the DFT and UHF calculations we must use suitable basis sets. Basis sets for atoms beyond the third row of the periodic table are handled somewhat differently. For these very large nuclei, electrons near the nucleus are treated in an approximate way, via effective core potential (ECPs). This treatment includes some relativistic effects, which are important in these atoms. The LANL2DZ basis set is the best known of these which was used by us for simulation[18]. The result of simulation after considering the relativistic effects is shown in table 1. By comparing the results in table 1, it is concluded that by adding relativistic effects we could got better result [19].

Also we know the unit cells of the ternary alloys are defined by Vegard's law[20] as follows:

$$a_{AB_xC_{1-x}} = (x)a_{AB} + (1-x)a_{AC}$$
$$a_{A_{1-x}B_xC} = (1-x)a_{AC} + (x)a_{BC} \qquad (3.1)$$

where $a_{AB_xC_{1-x}}$, $a_{A_{1-x}B_xC}$, $a_{AC}$, $a_{BC}$, and $a_{AB}$ are the lattice constant of the alloys and compounds respectively and InSb has zinc belende structure. If we assumed the Vegard's law is valid for nano clusters and they have zinc belende structure the We can consider $In_{10}Sb_4H_{23}$ cluster in zinc belende structure i.e its lattice constant equal to 6.4598 A. Since the tetragonal symmetry of InBi suggested that the grown InSbBi might have a tetragonal rather than zinc blende structure[21,22], we kept the dimensions in X-Y plane fix, and changed the dimension in Z direction (called h) after substituting Bi atoms instead of Sb atoms in the cluster. By using B3LYP/LANL2DZ Code, we calculated band gap, cohesive energy, nuclear repulsion energy and symmetry of cluster in density functional method based on the relativistic bases.

The band gap was estimated by the difference in the energies of the highest occupied and lowest unoccupied Khon-Sham eigenvalues.

The result is shown in table 2. As the results show, when h/4=1.62495, band gap of $In_{10}Sb_3BiH_{23}$ and $In_{10}Sb_2Bi_2H_{23}$ are smaller than $In_{10}Sb_4H_{23}$ (see item 3, 8 and 9). It means that, after adding the Bi to InSb, the structure was changed from zinc blende to tetragonal and band gap was decreased although the symmetry was not changed [23].

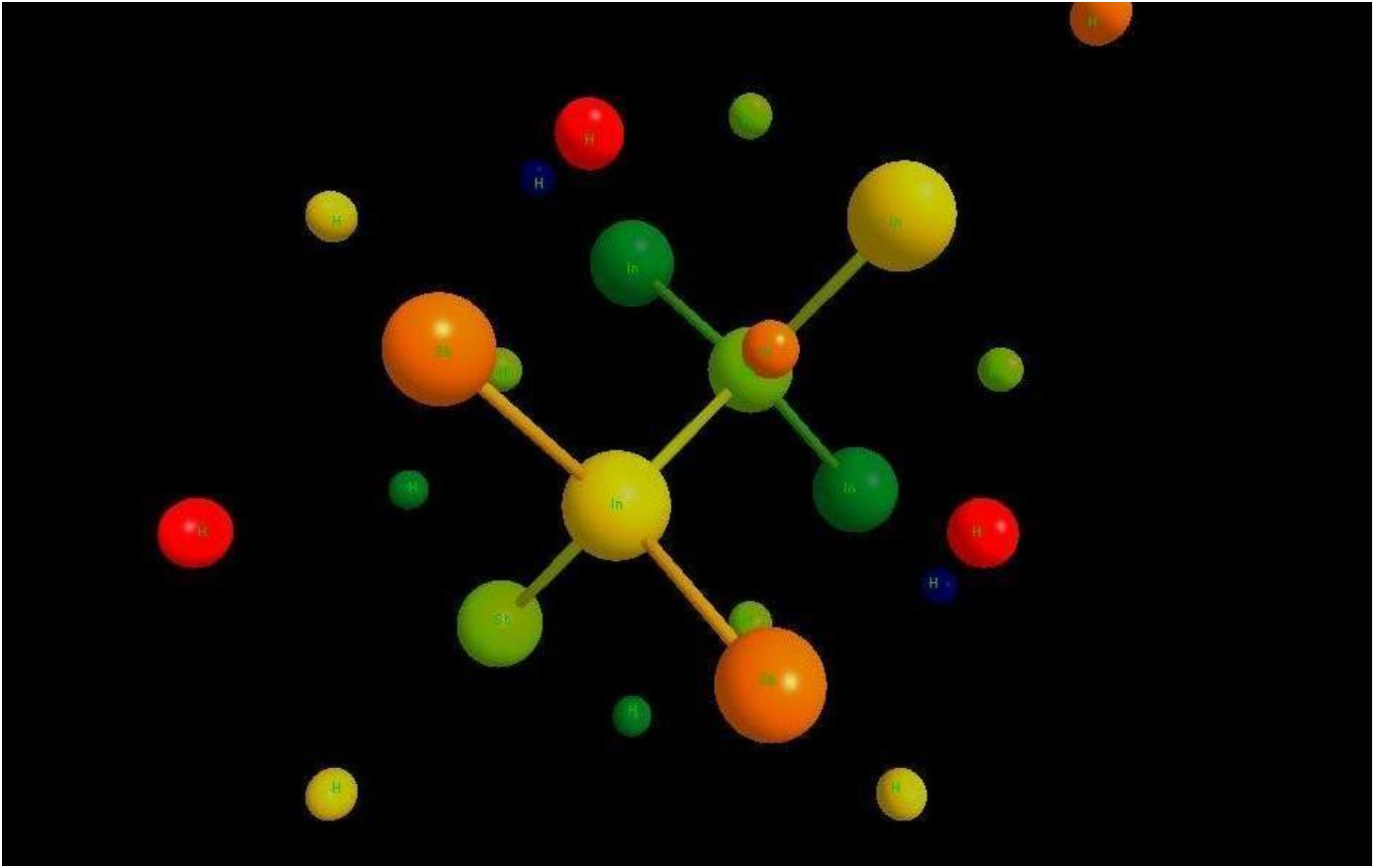

Figure 2. The In$_4$Sb$_4$H$_{18}$ Cluster

Table 2. The simulation results of In$_{10}$Sb$_4$H$_{23}$ before and after adding Bi element

| Item | Formula | Band Gap, eV | Cohesive energy, a.u | Nuclear Repulsion Energy, Hart. | Full point group | Largest Abelian group | Largest concise Abelian group | Lattice constant in Z direction, h/4, A |
|---|---|---|---|---|---|---|---|---|
| 1 | In$_{10}$Sb$_4$H$_{23}$ | 0.571 | -52.69 | 251.0587 | C1, NOP=1 | C1, NOP=1 | C1, NOP=1 | 1.61495 |
| 2 | In$_{10}$Sb$_3$BiH$_{23}$ | 0.557 | -52.75 | 251.0587 | C1, NOP=1 | C1, NOP=1 | C1, NOP=1 | 1.61495 |
| 3 | In$_{10}$Sb$_3$BiH$_{23}$ | 0.375 | -52.70 | 250.5582 | C1, NOP=1 | C1, NOP=1 | C1, NOP=1 | 1.62495 |
| 4 | In$_{10}$Sb$_3$BiH$_{23}$ | 0.507 | -52.73 | 250.05925 | C1, NOP=1 | C1, NOP=1 | C1, NOP=1 | 1.63495 |
| 5 | In$_{10}$Sb$_3$BiH$_{23}$ | 0.535 | -52.74 | 250.3085 | C1, NOP=1 | C1, NOP=1 | C1, NOP=1 | 1.62995 |
| 6 | In$_{10}$Sb$_3$BiH$_{23}$ | 0.572 | -52.73 | 250.8082 | C1, NOP=1 | C1, NOP=1 | C1, NOP=1 | 1.61995 |
| 7 | In$_{10}$Sb$_3$BiH$_{23}$ | 0.406 | -52.71 | 250.5557 | C1, NOP=1 | C1, NOP=1 | C1, NOP=1 | 1.625 |
| 8 | In$_{10}$Sb$_3$BiH$_{23}$ | 0.327 | -52.70 | 250.5607 | C1, NOP=1 | C1, NOP=1 | C1, NOP=1 | 1.6249 |

# 4 InAs Clusters

We considered $In_{19}Sb_{16}H_{36}$ and $In_{19}As_{16}H_{36}$ clusters in zinc belende structure i.e its lattice constants are equal to 6.4598 A and 6.058 A respectively.

After adding the As atoms to InSb or adding Sb atoms to InAs the lattice constant can be calculated by Vegard's law[20]. We considered x value equal to the number of guest atoms to the number of host atoms and used the calculated lattice constant by Vegard's law in the simulations. When the number of guest atoms be equal to the number of host atoms, we put the average lattice constant in the simulation.

Finally by using B3LYP/LANL2DZ Code, we calculated band gap, cohesive energy, nuclear repulsion energy and symmetry of cluster in density functional method based on the relativistic bases. The results are shown in tables 3 and 4.

As the table 3 shows, when the symmetry of the cluster, $In_{19}Sb_{16}H_{36}$, after adding the As element approaches to the symmetry of the cluster before adding the guest element, the band gap reduction is seen. It means that, not only the element but also the final symmetry of the cluster is important (see item 9 of table 3).

As table 4 shows, when the symmetry of the cluster, $In_{19}As_{16}H_{36}$, after adding the Sb element approaches to the symmetry of the cluster before adding the guest element, the band gap reduction is seen. It means that, not only the element but also the final symmetry of the cluster is important (see items 5 and 9 of table 4).

As the items 3 and 8 of table 2 show, when x value is equal to 1/3, the minimum band gap was seen. Also as the item 5 of the table 4 shows again when x value is equal to 4/12, the minimum band gap was seen. In the other words both of them at same value of x show minimum band gap[23].

Table 3. The simulation results of $In_{19}Sb_{16}H_{36}$ before and after adding As element

| Item | Formula | Band Gap, eV | Cohesive energy, a.u | Nuclear Repulsion Energy, Hart. | Full point group | Largest Abelian group | Largest concise Abelian group | Lattice constant, a/4, A |
|---|---|---|---|---|---|---|---|---|
| 1 | $In_{19}Sb_{16}H_{36}$ | 0.387 | -141.616 | 1191.2338 | TD, NOP=24 | D2, NOP=4 | D2, NOP=4 | 1.61495 |
| 2 | $In_{19}Sb_{15}AsH_{36}$ | 0.525 | -142.356 | 1196.0745 | C1, NOP=1 | C1, NOP=1 | C1, NOP=1 | 1.6082 |
| 3 | $In_{19}Sb_{14}As_2H_{36}$ | 0.502 | -143.091 | 1201.85086 | C1, NOP=1 | C1, NOP=1 | C1, NOP=1 | 1.6006 |
| 4 | $In_{19}Sb_{13}As_3H_{36}$ | 0.499 | -143.802 | 1208.5963 | C3V, NOP=6 | CS, NOP=2 | CS, NOP=2 | 1.59175 |
| 5 | $In_{19}Sb_{12}As_4H_{36}$ | 0.548 | -144.542 | 1216.2795 | C1, NOP=1 | C1, NOP=1 | C1, NOP=1 | 1.58145 |
| 6 | $In_{19}Sb_{11}As_5H_{36}$ | 0.489 | -145.264 | 1225.880 | C1, NOP=1 | C1, NOP=1 | C1, NOP=1 | 1.5693 |
| 7 | $In_{19}Sb_{10}As_6H_{36}$ | 0.717 | -146.001 | 1236.9331 | C1, NOP=1 | C1, NOP=1 | C1, NOP=1 | 1.5546 |
| 8 | $In_{19}Sb_9As_7H_{36}$ | 0.638 | -146.714 | 1251.7066 | C1, NOP=1 | C1, NOP=1 | C1, NOP=1 | 1.5368 |
| 9 | $In_{19}Sb_8As_8H_{36}$ | 0.257 | -147.428 | 1229.4901 | CS, NOP=2 | CS, NOP=2 | CS, NOP=2 | 1.5647 |

Table 4. The simulation results of $In_{19}As_{16}H_{36}$ before and after adding Sb element

| Item | Formula | Band Gap, eV | Cohesive energy, a.u | Nuclear Repulsion Energy, Hart. | Full point group | Largest Abelian group | Largest concise Abelian group | Lattice constant, a/4, A |
|---|---|---|---|---|---|---|---|---|
| 1 | $In_{19}As_{16}H_{36}$ | 0.969 | -153.270 | 1270.2430 | TD, NOP=24 | D2, NOP=4 | D2, NOP=4 | 1.5145 |
| 2 | $In_{19}As_{15}SbH_{36}$ | 0.488 | -152.512 | 1264.43 | C1, NOP=1 | C1, NOP=1 | C1, NOP=1 | 1.5212 |
| 3 | $In_{19}As_{14}Sb_2H_{36}$ | 0.405 | -151.738 | 1258.3615 | C2V, NOP=4 | C2V, NOP=4 | C2V, NOP=4 | 1.5288 |
| 4 | $In_{19}As_{13}Sb_3H_{36}$ | 0.834 | -151.009 | 1251.1597 | C3V, NOP=6 | CS, NOP=2 | CS, NOP=2 | 1.5376 |
| 5 | $In_{19}As_{12}Sb_4H_{36}$ | 0.229 | -150.214 | 1242.7539 | TD, NOP=24 | D2, NOP=4 | D2, NOP=4 | 1.548 |
| 6 | $In_{19}As_{11}Sb_5H_{36}$ | 0.545 | -149.518 | 1233.0628 | C1, NOP=1 | C1, NOP=1 | C1, NOP=1 | 1.5602 |
| 7 | $In_{19}As_{10}Sb_6H_{36}$ | 0.558 | -148.789 | 1222.0399 | C1, NOP=1 | C1, NOP=1 | C1, NOP=1 | 1.57477 |
| 8 | $In_{19}As_9Sb_7H_{36}$ | 0.509 | -148.079 | 1207.9512 | C3V, NOP=6 | CS, NOP=2 | CS, NOP=2 | 1.5926 |
| 9 | $In_{19}As_8Sb_8H_{36}$ | 0.381 | -147.323 | 1229.4901 | CS, NOP=2 | CS, NOP=2 | CS, NOP=2 | 1.5647 |

## 5  GaAs

we considered $Ga_4As_4H_{18}$, $Ga_{19}As_{28}H_{24}$, $Ga_4As_3BiH_{18}$ and $Ga_{19}As_{27}BiH_{24}$ clusters and calculated: cohesive energy, band gap by unrestricted Hartee-Fock (UHF) and density functional theory (DFT) methods.

The result of simulation of energy is shown in table 1. We considered the clusters have exactly the zinc bland structure of GaAs.

Table 5. Cohesive energy and Band gap of $Ga_4As_4H_{18}$ and $Ga_{19}As_{28}H_{24}$

|  | Cohesive energy (a.u) | | Band gap (ev) | |
|---|---|---|---|---|
|  | UHF | DFT | UHF | DFT |
| $Ga_4As_4H_{18}$ | 0.7369 | 0.8699 | 2.2198 | 1.1246 |
| $Ga_{19}As_{28}H_{24}$ | 1.3573 | 1.5931 | 5.5417 | 0.5366 |

The effect of band gap narrowing after adding Bi to $Ga_4As_3BiH_{18}$ and $Ga_{19}As_{27}BiH_{24}$ clusters and the amount of calculated band gap of $Ga_4As_3BiH_{18}$ show is shown in table 6. the LANAL2DZ basis was used.

Table 6. Cohesive energy and Band gap of $Ga4As_3BiH_{18}$ and $Ga_{19}As_{27}BiH_{24}$

|  | Cohesive energy (a.u) | | Band gap (ev) | |
|---|---|---|---|---|
|  | UHF | DFT | UHF | DFT |
| $Ga_4As_3BiH_{18}$ | 0.1812 | 0.2086 | 4.4818 | 1.0962 |
| $Ga_{19}As_{27}BiH_{24}$ | 0.7732 | 0.7247 | 2.6712 | 0.4753 |

The results of table 5 and table 6 show by inserting Bi instead of As in the both clusters the band gap decrease. As the result of UHF simulation on $Ga_4As_3BiH_{18}$ shows the band gap reach to ~1 ev after inserting Bi.

The ~1 ev value for band gap of GaAsBi alloys was seen theoretically and experimentally[24,25], but although the cluster size (i.e $Ga_4As_3BiH_{18}$) is small and the inclusion of the displacements of the nearest neighbor atoms did not considered, the result of UHF/STO-3G code shows bulk value.

## 6  CdSe

It was shown, in contrast to bulk CdSe, the nanosized samples is observed to retain the wurtzite structure well above the bulk phase transition pressure of 3 GPa upstroke and at pressure above 6 GPa, the sample begins to convert to the rock salt structure [26]. Also it was shown, the Alkoxysilane-stabilized CdSe nanocrystals exhibited cubic zinc blend structure [27,28].

In this reason we considered the zinc blend and wurtzite CdSe clusters for ab initio calculations.

We considered $CdSe_4H_{12}$ and $Cd_4Se_4H_{16}$ cluster in zinc-blend structure with lattice constant equal to 4.7 A. The clusters are shown in figure 3 and 4.

Also we considered $Cd_9Se_6H_8$ and $Cd_{15}Se_{10}H_{13}$ clusters in wurtzite structure with lattice constant equal to a=3.43816A and C=5.60168A. The clusters are shown in figure 5 and 6. The calculation results are shown in table 7, table 8, table 9 and table 10 [29].

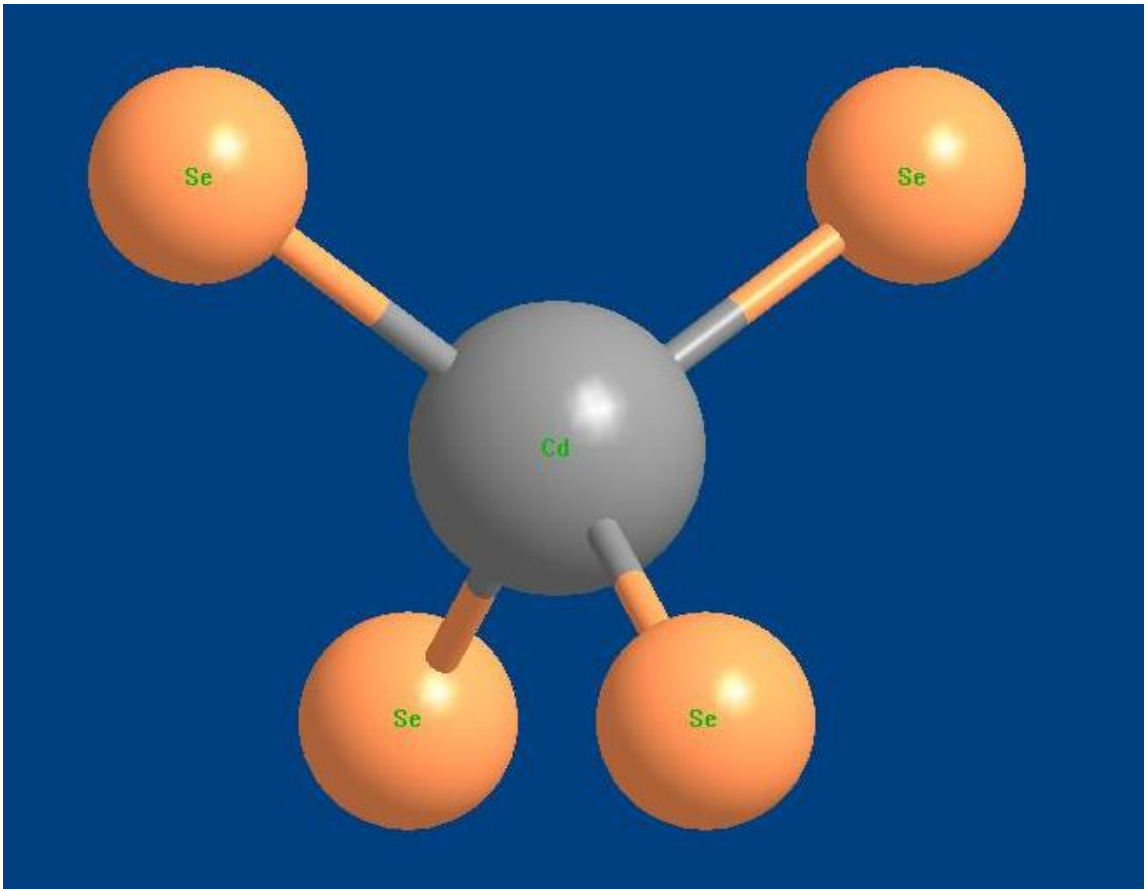

**Figure 3- Zinc-blend Structure of CdSe₄ cluster**

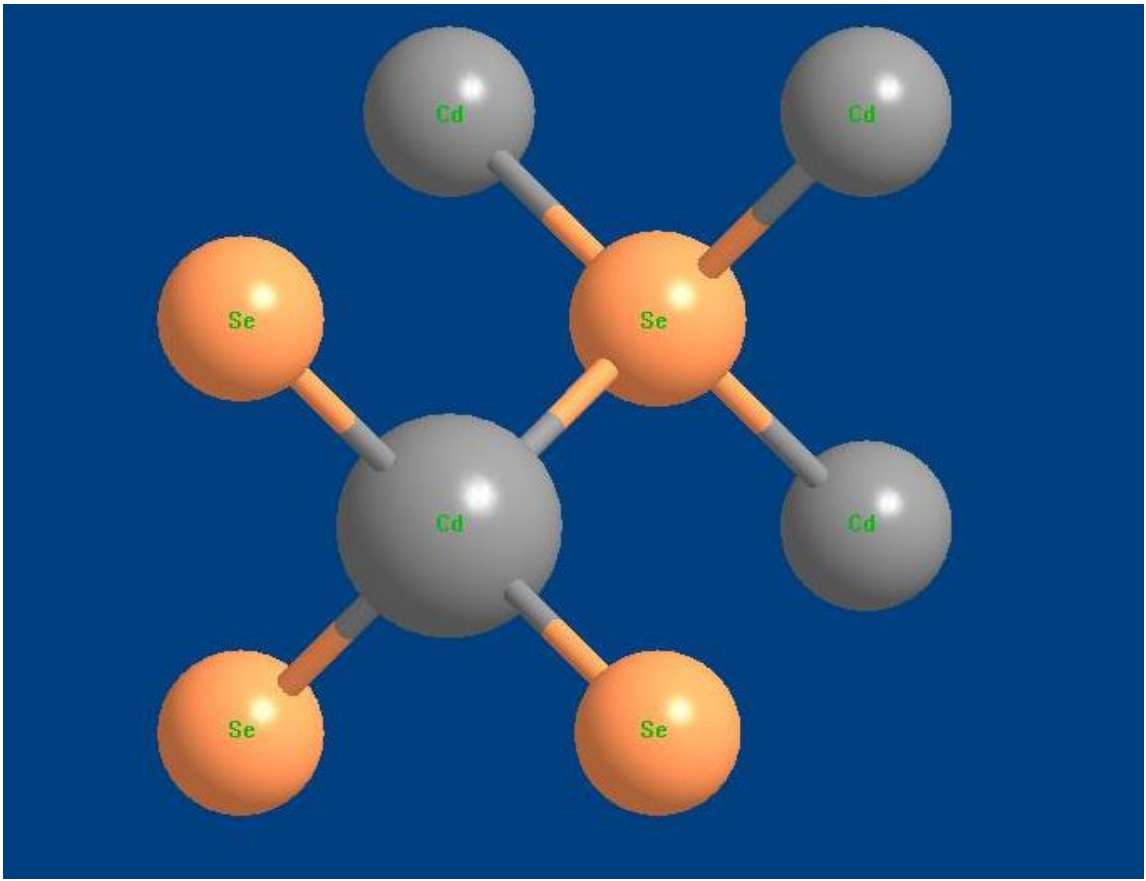

**Figure 4-Zinc-blend structure of Cd₄Se₄ cluster**

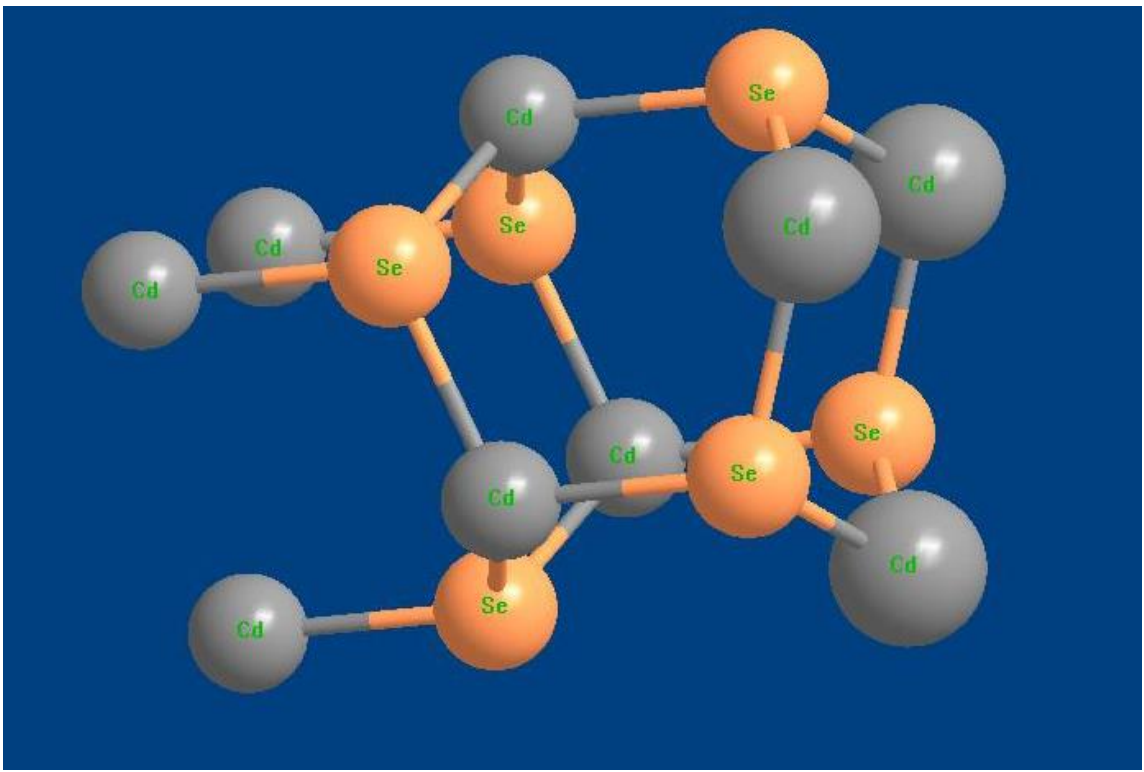

**Figure 5-Wurtzite Structure of Cd$_9$Se$_6$ clusters**

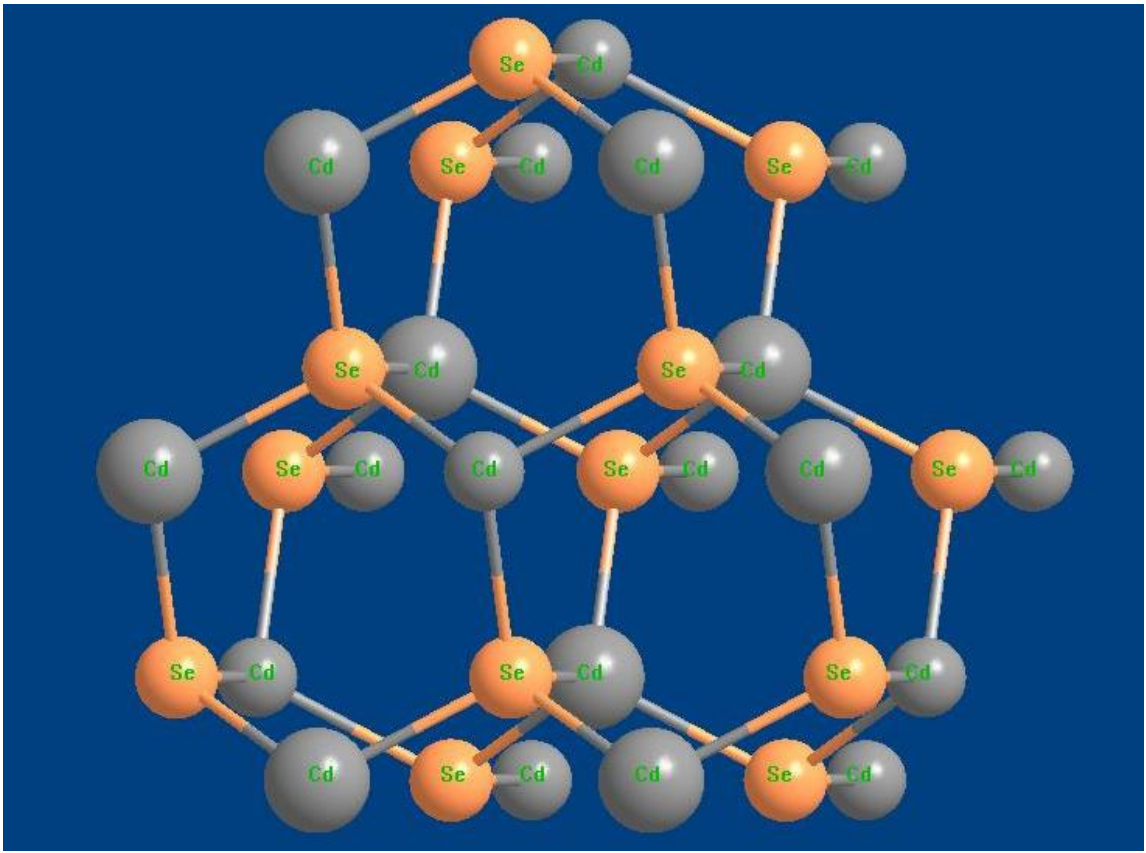

**Figure 6-Wurtzite Structure of $Cd_{20}Se_{13}$ cluster**

| Cluster Zinc-Blende | Cohesive energy, a.u | Band gap, eV |
|---|---|---|
| $CdSe_4H_{12}$ | 0.408 | 6.50 |
| $Cd_4Se_4H_{16}$ | 0.234 | 5.36 |

**Table7-The cohesive energy, band gap, of zinc-blend structure, Hartree-Fock**

| Cluster Zinc-Blende | Cohésive energy, a.u | Band gap, eV |
|---|---|---|
| $CdSe_4H_{12}$ | 0.387 | 2.86 |
| $Cd_4Se_4H_{16}$ | 0.243 | 1.59 |

**Table8-The cohesive energy, band gap, of zinc-blend structure, Density Functional Theory**

| Cluster Wurtzite | Cohésive energy, a.u | Band gap, eV |
|---|---|---|
| $Cd_9Se_6H_8$ | 0.092 | 5.31 |
| $Cd_{15}Se_{10}H_{13}$ | 0.007 | 3.56 |

**Table 9-The cohesive energy, band gap, of wurtzite structure, Hartree-Fock**

| Cluster Wurtzite | Cohésive energy, a.u | Band gap, eV |
|---|---|---|
| $Cd_9Se_6H_8$ | 0.12 | 1.18 |
| $Cd_{15}Se_{10}H_{13}$ | 0.036 | 1.37 |

**Table 10-The cohesive energy, band gap, of wurtzite structure, Density Functional Theory**